\documentclass[dvips]{article}

\usepackage{icrc2011}

\title{Remarks on the chemical composition of highest-energy cosmic rays }

\newcommand{\etal}{\MakeLowercase{\textit{et al. }}} 
\shorttitle{Wilk \etal chemical composition from eas }

\authors{Grzegorz Wilk$^{1}$, Zbigniew W\l odarczyk$^{2}$  }
\afiliations{$^1$The Andrzej So\l tan Institute for Nuclear Studies, Hoza 69,
00-681 Warsaw, Poland \\ $^2$ Institute of Physics, Jan Kochanowski University, \'Swi\c{e}tokrzyska 15,
 25-406 Kielce, Poland }
\email{Zbigniew.Wlodarczyk@ujk.kielce.pl}

\abstract{We present arguments aiming to reconcile the apparently
contradictory results concerning the chemical composition of
cosmic rays of highest energy, coming recently from Auger and
HiRes collaborations. In particular, we argue that the energy
dependence of the mean value and root mean square fluctuation of
shower maxima distributions observed by the Auger experiment are
not necessarily caused by the change of nuclear composition of
primary cosmic rays.}

\begin{document}
\maketitle

\section{Introduction}

The identities of highest-energy cosmic rays remains still an
open question. Possible conclusions on either protons or
iron nuclei dominance in cosmic ray flux leads to problems
\cite{BS_rev}. Seeking to determine the nuclear identities of
ultrahigh-energy cosmic-ray (UHECR) particles, the development of
extensive air showers (EAS) of secondary particles in the
atmosphere was extensively examined. The Auger collaboration
\cite{Auger} has determined both the shower maximum $\langle
X_{max}\rangle $ (the penetration depth in the atmosphere at which
the shower reaches its maximum number of secondary particles) 
and the complementary observable $ \sigma(X_{max}) $  (the root
mean square fluctuation of $ X_{max}$  from event to event). Their
results seem to indicate a transition, at primary energies of a
few times $10^{18} $ eV, from the flux dominated by protons
to the one increasingly dominated at higher energies by iron
nuclei. The HiRes collaboration \cite{HiRes} has analyzed
event-by-event fluctuations of  data in terms of the
truncated fluctuation widths  $ \sigma_{T} $ (  the $ X_{max}
$ distribution was truncated at $ 2\sigma (X_{max})$ ), and
reaches a different conclusion. We would like to present here
arguments that the observed energy dependence of $ \left\langle
X_{max}\right\rangle $ and $ \sigma(X_{max}) $ by Auger experiment
are not originated by the changes of nuclear composition of
primary cosmic rays (cf., also, \cite{WW}) and that the
highest-energy cosmic rays seems to be dominated by protons.

\section{Inconsistency in the iron abundance}

 \begin{figure}[!t]
  \vspace{5mm}
  \centering
  \includegraphics[width=2.5in]{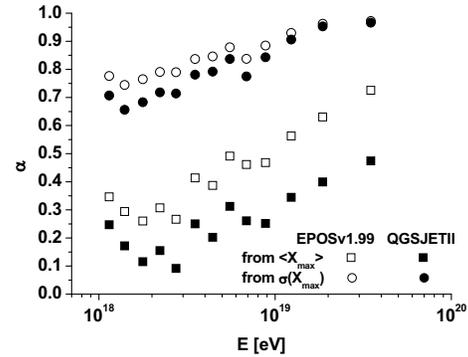}
  \caption{ The energy dependence of relative abundance of iron in CR as
  extracted from $ \langle X_{max}\rangle $  and $ \sigma(X_{max}) $ given
  by Auger experiment \cite{Auger} (in frame of QGSJETII \cite{QGSJETII}
  and EPOSv1.99 \cite{EPOS} models). }
  \label{fig1}
 \end{figure}

With the energy increase, the spectacular Auger data \cite{Auger}
show almost monotonic changes from proton composition towards iron
one for both  $ \langle X_{max}\rangle $ and $ \sigma(X_{max}) $
observables. For  $ \langle X_{max}\rangle $ such dependence 
can be easily interpreted by two component cosmic ray composition
(with relative abundance of iron nuclei $ \alpha $ and
contribution of protons $ 1-\alpha $ ), for which we expect
 \begin{equation}
\left\langle X_{max}\right\rangle =\left( 1-\alpha\right) \left\langle X_{max}\right\rangle _{p}
+\alpha \left\langle X_{max}\right\rangle _{Fe} \label{eq:Xmax}
\end{equation}
where $ \langle X_{max}\rangle _{p} $  and $ \langle
X_{max}\rangle _{Fe}$  are the shower maxima for pure proton and
iron nuclei, respectively. However for $ \sigma(X_{max}) $ we have
nonmonotonic dependence on  $ \alpha $,
\begin{eqnarray}
 \sigma^{2}=\left( 1-\alpha\right) \sigma_{p}^{2}+\alpha\sigma_{Fe}^{2}+ \nonumber \\
 +\alpha\left( 1-\alpha\right)
 \left( \left\langle X_{max}\right\rangle _{p}-\left\langle X_{max}\right\rangle _{Fe}\right)
 ^{2}.
\label{eq:RMS}
\end{eqnarray}
For this reason the experimental data  (with similar energy
behavior) lead to quite different chemical composition, ranging
from the proton dominated for $ \langle X_{max}\rangle $ to  the iron
dominated for $ \sigma(X_{max}) $ (cf. Fig.1).

\section{Importance of the first interaction point}

Some remarks are in order at this point (cf. \cite{Alvarez, Risse,
Urlich, Alvarez2}). Most of the charged particles in the shower
are electrons and positrons coming from the electromagnetic
subshowers initiated by photons from  $ \pi^{0} $-decay, with
energies near the critical energy ($ \varepsilon = 81$ MeV in
air). The mean depth of maximum for an electromagnetic shower
initiated by a photon with energy $ E_{\gamma} $  is
\begin{equation}
\left\langle X_{max}^{em}\left( E_{\gamma}\right) \right\rangle
=X_{0}\ln \left( E_{\gamma}/\varepsilon\right), \label{eq:X_em}
\end{equation}
where  $ X_{0}\approx $37 $g/cm^{2}$ is the radiation length in
air. A nuclear-initiated shower consists of a hadronic core
feeding the electromagnetic component primarily through $ \pi^{0}
$ production. In general, for an incident nucleus of mass $ A $
and total energy $ E $ (including protons with $ A $=1) the depth
of maximum is expressed by
\begin{equation}
\left\langle X_{max}\left( E\right) \right\rangle = \left\langle
X_{max}^{em}\left( \left( E/A\right) \left( K/\left\langle
n\right\rangle \right) \right) \right\rangle +\left\langle
X_{1}\right\rangle, \label{eq:X_max}
\end{equation}
where $ \left\langle X_{1}\right\rangle  $  is the mean depth of
the interaction with maximal energy deposition into shower
(usually called the depth of the first interaction),  $ K $
denote inelasticity and $ \left\langle n\right\rangle  $  is
related to the multiplicity of secondaries in the high-energy
hadronic interactions in the cascade. If the composition changes
with energy, then $ \left\langle A\right\rangle  $ depends on
energy and $ \left\langle X_{max}\right\rangle  $   changes
accordingly. The situation is, however, essentially more
complicated. Whereas  for a primary nucleus in which the energy is
to a good approximation simply divided into $ A $ equal parts, in
a hadronic cascade there is instead a hierarchy of energies
of secondary particles in each interaction, and a similar
(approximately geometric) hierarchy of interaction energies in the
cascade. In this case $ \left\langle n\right\rangle $ has to be
understood as some kind of "effective" multiplicity, which does
not have a straightforward definition in general. For this reason
the change of primary composition or the violation of Feynman
scaling are widely discussed since many years. In addition
to this, the inelasticity $  \left\langle K\right\rangle $ can
itself be function of energy \cite{K}.

The probability of having the first interaction point of a
shower, $ X_{1} $  , at a depth greater than $ X $ is
\begin{equation}
P(X_{1}>X)\sim\exp \left( -X/\lambda\right), \label{eq:p1}
\end{equation}
where $ \lambda $ is the interaction length. In the
case of perfect correlation between $ X_{max} $ and  $ X_{1} $,
i.e., when fluctuations in the shower development were
nonexistent, one could use directly the exponential distribution
of showers with large $ X_{max} $  to calculate $ X_{1}$  and
hence the proton-air cross section. However, intrinsic shower
fluctuations modify relation between the depth of maximum
distribution and the interaction length. This modification is
typically expressed by a factor $ k=\Lambda/\lambda $  and
leads to $ P(X_{max}>X)\sim\exp (-X/\Lambda) \label{p2}$. The
factor $ k $ depends mainly on how fast is the energy
dissipation in the early stages of shower evolution. In
particular it is sensitive to the mean inelasticity and to
its fluctuations. In general, a model with small fluctuations in
secondary particle multiplicity and inelasticity is characterized
by a smaller $ k $ factor than a model with large fluctuations.
Under the assumption of similar fluctuations in multiplicity and
inelasticity, a model predicting a large average number of
secondary particles leads to smaller overall fluctuations of the
cumulative shower profile of the secondary particles and hence to
a smaller $ k $ factor.

In the absence of internal fluctuations, all showers would develop
through the same amount of matter, $ \Delta X=X_{max}-X_{1}$,
between the first interaction point and the maximum. As a
consequence, a perfect correlation between  $ X_{max} $ and $
X_{1} $  would exist, and their distributions would have exactly
the same shape, shifted by a constant  $ \Delta X $. In that case
the slope of the $ X_{max} $   distribution, $ \Lambda $,  would
be equal to the mean interaction length, $ \lambda $. Intrinsic
fluctuations in shower development (after the first interaction)
affect the relation between the interaction length  $ \lambda $
and the slope $ \Lambda $ that describes the exponential tail of
the $ X_{max} $ distribution. The relation is often expressed with
a $ k $ factor $ k=\Lambda/\lambda$. For more properties of EAS
and influence of shower fluctuations on studies of the shower
longitudinal development see Ref.
\cite{Alvarez,Risse,Urlich,Urlich2,Alvarez2}.

The effect of fluctuations  in $ \Delta X $  is to broaden the
correlation  of $ X_{max} $ with $ X_{1}$. However, we can
roughly  write that
\begin{equation}
\sigma(X_{max})\cong \sigma(X_{1})+\xi \left( \sigma(\Delta
X)\right), \label{eq:sig}
\end{equation}
where  $ \sigma (X_{1}) \propto <X_{1}> $ and the function $
\xi $ describes influence of shower fluctuations after the first
(main) interaction point (notice that for the probability
distribution given by Eq.( \ref{eq:p1}) the fluctuation in $
X_{1} $ is $\sigma(X_{1})=\sqrt{Var(X_{1})}=\left\langle X_{1}\right\rangle$, 
whereas for $ X_{1} $ interpreted as the main interaction point we have
$\sigma(X_{1})=\left\langle X_{1}\right\rangle/\sqrt{\kappa} $ where $ \kappa$ 
determines in which of the succesive interactions of projectile particle 
the energy deposition to the shower is maximal). Because of Eq.(\ref{eq:X_max}), 
where $ \left\langle X_{max}\right\rangle
=\left\langle X_{max}^{em}\right\rangle  + \left\langle
X_{1}\right\rangle$, we can construct observable in which
influence of fluctuation of the first interaction point is strongly
suppressed, namely
\begin{eqnarray}
&&\left\langle X_{max}\right\rangle -\sigma (X_{max}) \cong  \nonumber \\
&&\cong \left\langle X_{max}^{em}\left((E/A)(K/<n>\right)
\right\rangle + \xi\left( \sigma(\Delta X)\right).
 \label{eq:x_sig}
\end{eqnarray}

\section{Results}

In Fig.2 this observable is plotted for Auger \cite{Auger} and
HiRes \cite{HiRes} data in comparison with different models
\cite{QGSJETII, EPOS, QGSJET01, Sibil}. To make the results
from both experiment to coincide, the HiRes data are shifted by
$10$ $  g/cm^{2} $ (in this case predictions from QGSJETII model are
roughly the same for both experiment) \footnote{ Notice that
Auger compares the data with pure simulations. HiRes quotes data
including all detectors effect and compare it to the models
'after' the detector simulation. Unfortunately that means that
both approaches cannot be compared directly.}.

 \begin{figure*}[th]
  \centering
  \includegraphics[width=5in,height=3.2in]{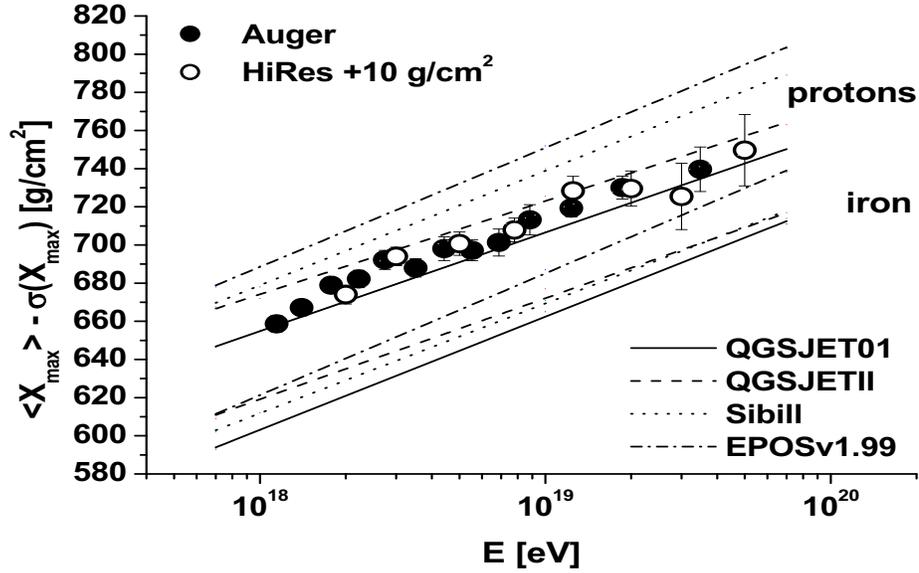}
  \caption{ $ \left\langle X_{max}\right\rangle -\sigma(X_{max}) $
  as deduced from Auger data  and
  $ \left\langle X_{max}\right\rangle -\sigma_{T}(X_{max}) $
  from HiRes data in comparison with different models.
  Notice that HiRes experimental data are shifted by $10$ $g/cm^{2}$
  to make the model predictions for both experiment coincide.
    }
  \label{fig.2}
 \end{figure*}

Notice that  $ \left\langle X_{max}\right\rangle -\sigma(X_{max})
$ is still dependent on models and, in particular,  it is
sensitive to the chemical composition. Showers initiated by
protons are seemingly different from those initiated by iron
nuclei.

From Fig.2 we can learn that the chemical composition is not the
origin of the effect observed by Auger experiment. Moreover,
the experimental data fairly well coincide with the proton
dominant primary composition. Within the toy model of primary
composition (only two components: iron nuclei with relative
abundance $ \alpha $ and protons with abundance $ 1-\alpha $ ) we
can evaluate  $ \alpha $ from  $ \left\langle X_{max}\right\rangle
-\sigma(X_{max}) $ as given by Auger experiment. The results is
shown in Fig.3. For the reference model QGSJETII the
abundance of iron is roughly independent on energy ($
\alpha\simeq 0.05\div 0.1 $) and even for model EPOS v.1.99
\cite{EPOS},  which leades to the maximal abundance of iron,
it increases slowly with energy (varying in interval $
\alpha\simeq 0.15\div 0.3 $). The iron abundance shown in Fig.3
coincides with the one which can be estimated from HiRes
data. The comparison of $\alpha$ from Auger and HiRes data is
shown in Fig.4. In the energy region $2\cdot 10^{18}\div
5\cdot 10^{19}$ eV the mean  values of $\alpha$, evaluated
from $ \left\langle X_{max}\right\rangle -\sigma(X_{max})$, are
equal $\alpha = 0.08 \pm 0.01 $ from Auger data and $\alpha = 0.06
\pm 0.05 $ from HiRes data (notice that HiRes data on $
\left\langle X_{max}\right\rangle$ result in comparable
value, $\alpha=0.03 \pm 0.02$).

 \begin{figure}[!t]
  \vspace{5mm}
  \centering
  \includegraphics[width=2.5in]{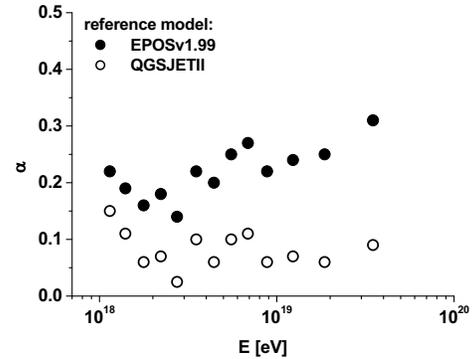}
  \caption{The energy dependence of relative abundance of iron
  in CR as extracted from $ \left\langle X_{max}\right\rangle -\sigma(X_{max}) $
  as given by Auger experiment and shown in Fig.2. }
  \label{fig3}
 \end{figure}

\section{Possible interpretation}

\begin{figure*}[th]
  \centering
  \includegraphics[width=5in,height=3.2in]{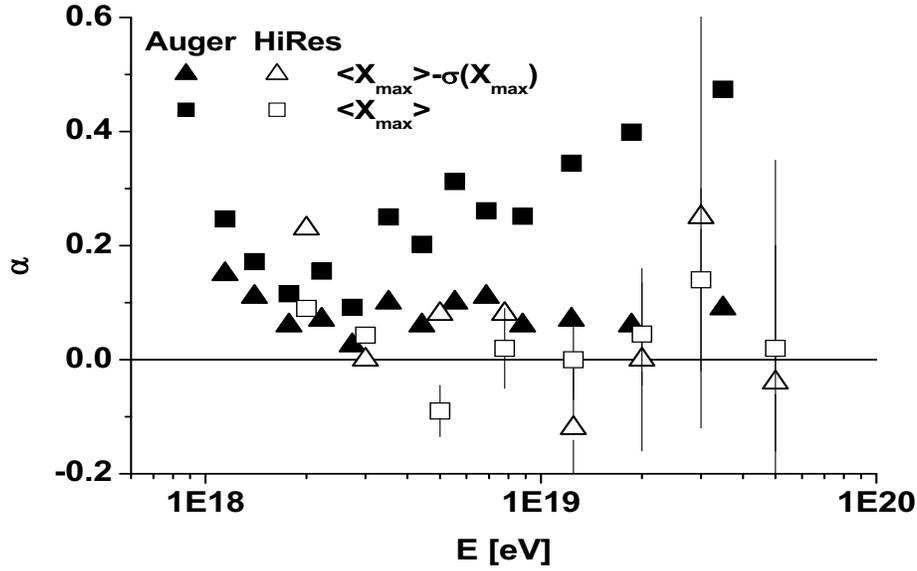}
  \caption{ The energy dependence of relative abundance of iron in
  CR as seen from Auger and HiRes data
   (in the frame of QGSJETII model \cite{QGSJETII}).
    }
  \label{fig.4}
 \end{figure*}

From Fig.2 we can learn that  $ \left\langle X_{1}(E)\right\rangle
$ gives the main contribution to the energy dependence of $
\left\langle X_{max}\right\rangle $ and $ \sigma (X_{max})$
observed experimentally. Two factors can affected energy
dependence of $ \left\langle X_{1}(E)\right\rangle$: the
cross section (interaction mean fee path $ \lambda $) and 
the inelasticity $ K $. Roughly, $ \left\langle X_{1}\right\rangle = \lambda\cdot\kappa
$, where $ \kappa $ determines in which of the successive
interactions of projectile the energy deposition to the shower is
maximal. For a uniform inelasticity distribution in the
maximal possible interval for a given $\left\langle K\right\rangle$ one has  $
\kappa\simeq 1+1.85(0.75-\left\langle K\right\rangle)$. The rapid increase of inelastic
cross section in energies $ E > 10^{18}$ eV cannot be
excluded. In particular, if gluon saturation occurs in the
nuclear surface region, the total cross section of proton$ -
$nucleus collisions increases more rapidly as a function of the
incident energy compared to that of a Glauber-type estimate
\cite{Portugal}. Although in \cite{K} the decrease of inelasticity
$ \left\langle K\right\rangle $ with energies was discussed in lower energy region, its
increase at energies $ E\sim 10^{18} $ eV is be no means excluded
(cf. the percolation effects which in high energies leads to
increase of inelasticity \cite{Dias}). Both possibilities are
questionable and require an abrupt onset of "new physics" beyond
the standard model (notice however that here, the center of mass
collision energy is about few hundreds of TeV, far beyond that can
be studied at LHC). Taking into account the HiRes data (where $
X_{max} $ distribution was truncated at $ 2\sigma $ ) we can learn
that the tails of $ X_{max} $  distribution are crucial. For this
reason, the role of biases due to the small statistics in
analyzing CR data of highest energy remains an open question (cf.
ref. \cite{WW}). It is interesting to note that the observable $
\left\langle X_{max}\right\rangle -\sigma(X_{max}) $ is rather
insensitive to the possible biases of the tail of $X_{max}$
distribution \cite{WW}.

\section{Concluding remarks}

To summarize, we conclude that the spectacular energy dependence
of the shower maxima distribution reported by Auger collaboration
\cite{Auger} is not necessarily (or not only) due to the changes
of chemical composition of primary cosmic rays. The observed
effect seems rather to be caused by the unexpected changes of the
depth of first interaction in energies above $ 2\,10^{18} $ eV.
They would requires, however, an abrupt onset of some "new
physics" in this energy region and are therefore questionable. We
argue that it would be highly desirable to analyze the observable
$ \left\langle X_{max}\right\rangle -\sigma(X_{max}) $ in which
fluctuations of the depth of the first interaction, as well as the
possible biases of the tail of $X_{max}$ distribution, are
strongly suppressed. This observable still depends on the model of
multiparticle production and is sensitive to the chemical
composition of the primary cosmic rays.


\clearpage

\end{document}